\documentclass[a4paper,english,aps,pra,showpacs,twocolumn]{revtex4}
\usepackage[T1]{fontenc}
\usepackage[latin1]{inputenc}
\usepackage{prettyref}
\usepackage{graphicx}

\makeatletter

\renewcommand{\vec}[1]{\mathbf{#1}}

\usepackage{babel}
\makeatother
\begin{document}

\title{Non-universal BCS-BEC crossover in resonantly interacting Fermi gases}

\author{L. M. Jensen$^{1,2}$}

\address{\textit{$^{1}$NORDITA, Blegdamsvej 17, DK-2100, Copenhagen Ø, Denmark}\\
\textit{$^{2}$Department of Theoretical Physics, Umeå University,
901 87, Umeå, Sweden}}

\email{melwyn@nordita.dk}


\begin{abstract}
We investigate the non-universal behavior of the BCS-BEC crossover
model at the normal to superfluid transition. By using a finite temperature
quantum field theoretical approach due to Nozières and Schmitt-Rink
and by making the effective range expansion of the effective two-body
interaction we numerically calculate the crossover transition temperature
as a function of the scattering length, $a_{F},$ and the effective
range parameter, $r_{\mathrm{e}}.$ In an ultracold two-component
atomic Fermi gas BCS-BEC crossover physics is expected to appear near
magnetic-field-induced Feshbach resonances. By matching two-body scattering
properties near a Feshbach resonance to a simple renormalized model
potential, a broad resonance is characterized by a small effective
range and the gas displays a universal BCS-BEC crossover behavior.
On the other hand, for a narrow resonance thermodynamic quantities
depend the effective range which may be large and negative at resonance.
For increasing values of $-r_{\mathrm{e}}$ we find the transition
temperature to be suppressed in the crossover region. We furthermore
argue for existence of a lower bound of $T_{c}$ at fixed coupling
for increasing $-r_{\mathrm{e}}$ in the crossover region. 
\end{abstract}
\maketitle

\section{Introduction }

The prospect of creating a fermion superfluid in trapped atomic Fermi
gases using a magnetic field induced Feshbach resonance has received
much experimental and theoretical attention in the past year. Recently
long-lived diatomic molecules of fermionic $^{6}\mathrm{Li}$ and
$^{40}\mathrm{K}$ \cite{Regal2003a,Cubizolles2003,Strecker2003},
Bose-Einstein condensation (BEC) of molecular $^{6}\mathrm{Li}_{2}$
and $^{40}\mathrm{K}_{2}$ dimers \cite{Jochim2003b,Greiner2003b,Zwierlein2003b},
crossover from a molecular BEC to a degenerate and strongly interacting
Fermi gas \cite{Bartenstein2004a,Bartenstein2004b}, and condensation
of fermionic atom pairs \cite{Regal2004a,Zwierlein2004a} have been
reported. From measurements of the energy gaps in the excitation spectrum
\cite{Chin2004a,Kinnunen2004} and of the frequencies and damping
of compressional collective modes as a function of the effective atom-atom
interaction \cite{Kinast2004a,Bartenstein2004b} evidence is emerging
that some of the experiments are already probing the superfluid phase. 

From a theoretical point of view the low temperature atom gases with
resonant interactions have on the one hand been described in terms
of coupled boson-fermion models where the atomic and molecular degrees
of freedom have been treated separately but constrained by total fermion
number conservation. On the other hand, the description of superfluid
pairing in Fermi systems with a varying coupling constant has also
been analyzed using a generalized Bardeen-Cooper-Schrieffer (BCS)
variational function at zero temperature \cite{Leggett1980} and by
Green's function methods at nonzero temperatures \cite{Nozieres1985,Melo1993}
including explicitly only the fermionic degrees of freedom. In the
latter approach a boson-like two-body bound state is formed at large
couplings and the system displays a smooth transition from a gas of
fermions which undergoes a phase transition to a BCS superfluid to
a gas of bosons which condenses at the Bose-Einstein condensation
temperature. 

In \cite{Bruun2004a,Palo2004,Diener2004b} it was argued that the
universal crossover model based on a static contact potential applies
only to broad Feshbach resonances, and that for more general potentials
(\emph{e.g.,} potentials characterized by the presence of a repulsive
barrier) the solutions to the crossover equations depend sensitively
on the effective range, which gives a measure of the energy dependence
of the scattering phase shift and transition matrix. Similar conclusions
were drawn in \cite{Simonucci2004} based on a microscopic five coupled
channels calculation of diatomic $^{6}\mathrm{Li}.$ In this study
the molecular Born-Oppenheimer potentials were adjusted in order to
reproduce the experimental resonance positions of the broad resonance
at $822$ G, the narrow resonance at $543$G in addition to the zero
crossing of the scattering length at $533$ G and the hyperfine splitting
between $F=3/2$ and $F=1/2$ states. The effective range was extracted
and found to be small and positive for the broad resonance and thus
consistent with the standard BCS-BEC crossover model. In contrast,
the effective range for the narrow resonance was found to be large
and negative. In relation to the two-channel scattering model derived
from the boson-fermion models the relation between the effective range,
$r_{\mathrm{e}},$ at resonance and the boson-fermion coupling constant,
$g_{BF},$ was found to be, $r_{\mathrm{e}}=-8\pi\hbar^{4}/(m^{2}|g_{BF}|^{2}),$
where $m$ is the mass of the atoms and when neglecting the background
scattering length \cite{Bruun2004a}. 

In the present work we generalize the calculation of the BCS-BEC crossover
transition temperature to potential models with nonzero effective
range. The paper is organized as follows: Section \ref{sec:intro}
contains a description of the finite temperature Green's function
formalism \cite{Nozieres1985} due to Nozières and Schmitt-Rink (NS).
Partly following \cite{Haussmann1993} the thermodynamic equations
determining the transition temperature and chemical potential are
presented, and the renormalization of the T-matrix equation including
the effective range expansion of the effective interaction is explained.
In Section \ref{sec:transition-temperature} the resulting equations
are numerically solved to determine the critical chemical potential
and transition temperature in the crossover region as a functions
of the effective range.

\section{BCS-BEC crossover model \label{sec:intro}}

The BCS-BEC crossover problem has a long history \cite{Eagles1969b,Leggett1980,Nozieres1985,Melo1993}
and it was recently revived in the context of high-$T_{c}$ superconductivity
of the cuprates. The generic crossover problem has also found applications
in several other areas where pairing correlations are stronger than
in the BCS theory, \emph{e.g.}, the case of a strongly interacting
Fermi gas with large scattering length. Let us consider the grand
canonical Hamiltonian, $K=H-\mu N,$ of a two-component atomic Fermi
gas, \begin{eqnarray}
K & = & \sum_{\vec{k},\sigma}\xi_{\vec{k}}a_{\vec{k}\sigma}^{\dagger}a_{\vec{k}\sigma}\nonumber \\
 & + & \sum_{\vec{q},\vec{k},\vec{k}^{\prime}}\frac{U_{\vec{k}\vec{k}^{\prime}}}{\mathcal{V}}a_{\vec{k}+\frac{1}{2}\vec{q}\uparrow}^{\dagger}a_{-\vec{k}+\frac{1}{2}\vec{q}\downarrow}^{\dagger}a_{-\vec{k}^{\prime}+\frac{1}{2}\vec{q}\downarrow}a_{\vec{k}^{\prime}+\frac{1}{2}\vec{q}\uparrow},\end{eqnarray}
 where $a_{\vec{k}\sigma}^{\dagger},a_{\vec{k}\sigma}$ are the creation
and annihilation operators for free atoms with momentum $\hbar\vec{k},$
energy $\varepsilon_{\vec{k}}=\hbar^{2}k^{2}/2m,$ the magnitude of
the momentum $k=|\vec{k}|,$ with the pseudospin index $\sigma=\uparrow,\downarrow$
denoting one of the two trapped hyperfine spin states, and with $m$
and $\mu$ being the mass and chemical potential of the atoms, respectively.
The single atom kinetic energy measured with respect to the chemical
potential is $\xi_{\vec{k}}=\varepsilon_{\vec{k}}-\mu,$ and the momentum
dependent Fourier transform of the real space inter-atomic interaction
is $U_{\vec{k}\vec{k}^{\prime}}=U(|\vec{k}-\vec{k}^{\prime}|)$ which
in the crossover model with a bare attractive contact interaction
has the constant strength, $U_{\vec{k}\vec{k}^{\prime}}=U.$ Due to
the singular nature of the contact interaction unphysical divergences
arise in three dimensions. The divergences may be removed by a standard
regularization procedure \cite{Fetter1971}. By relating the bare
potential to the two-body scattering length $a_{F}$ and the corresponding
effective strength, $U_{\ast}=4\pi\hbar^{2}a_{F}/m,$ the bare potential
can be eliminated in favor of the scattering length. This is formally
accomplished by using the relation between bare and renormalized strengths,
$U^{-1}=U_{\ast}^{-1}-\frac{1}{\mathcal{V}}\sum_{|\vec{k}|<k_{c}}(2\varepsilon_{\vec{k}})^{-1},$
where $\mathcal{V}$ is the volume of the system, and $\hbar k_{c}$
is a momentum cutoff. At fixed $a_{F}$ it is a relation between $U$
and $k_{c}.$ The final results become finite and $k_{c}$ independent
by taking the limit $k_{c}\to\infty$ at the end of the calculation.
The $1/\varepsilon_{\vec{k}}$ term regulates the divergence of the
free two-particle propagator. A similar procedure has been applied
to the general case of model potentials with non-zero range and is
presented in detail in the next section.

\subsection{Nonzero temperature Dyson equations }

In the standard crossover model \cite{Eagles1969b,Leggett1980,Nozieres1985,Melo1993}
the BCS gap equation is supplemented by an equation for the number
density. At fixed particle density this is an equation for the chemical
potential, which in the BCS limit is the Fermi energy, $\mu\approx E_{F}.$
In the BEC limit it is half of the bound state energy, $\mu=-E_{0}/2,$
with $E_{0}=\hbar^{2}/ma_{F}^{2}$. In the nonzero temperature Green's
function formulation \cite{Nozieres1985} of the crossover model the
appearance of the bound states with finite center-of-mass momentum,
$\vec{q},$ at sufficiently large interaction strength is accounted
for by use of the T-matrix approximation for the self-energy. The
T-matrix approximation diagrammatically corresponds to a resummation
of an infinite series of particle-particle ladder diagrams. By using
methods of nonzero temperature Green's function theory the equations
\begin{figure}
\begin{center}\includegraphics[%
  width=8cm]{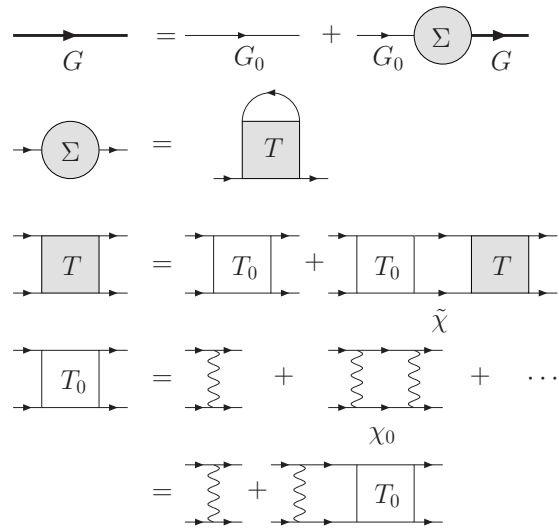}\end{center}

\caption{Diagrammatic structure of the equations for the dressed finite temperature
propagator $G$ (thick arrow lines). First line is the Dyson equation
for the atomic propagator, the second line is the self-energy (shaded
disc) derived from the T-matrix approximation, where the T-matrix
(shaded square) is found from the integral equation in the third line.
$T_{0}$ (white square) is the effective two-body interaction which
is found after proper renormalization of the bare two-body interaction
$U$ (wiggly lines) by inverting the equation for $T_{0}$ in the
fourth line. The non-interacting propagators are represented by the
thin arrow lines, and the internal pair of fermions in the third and
fourth lines are the regularized pair propagator, $\tilde{\chi},$
and non-interacting pair propagator, $\chi_{0},$ respectively. \label{cap:diagrams}}
\end{figure}
 for the transition temperature and critical chemical potential are
derived from the corresponding Dyson equation within the T-matrix
approximation,\begin{equation}
G(\vec{k},\omega_{n})=G_{0}(\vec{k},\omega_{n})+G_{0}(\vec{k},\omega_{n})\Sigma(\vec{k},\omega_{n})G(\vec{k},\omega_{n}),\label{eq:dyson}\end{equation}
with $\vec{k}$ being an atomic wave-vector, $\omega_{n}=(2n+1)\pi/\beta$
is the fermionic Matsubara frequency for integer $n$, and where $G_{0}$
and $G$ are the noninteracting and dressed finite temperature Green's
functions for the atoms, respectively (see Figure \ref{cap:diagrams}
for a diagrammatic representation). Unless specified we use below
$\hbar=1,$ $k_{B}=1,$ and $\beta=1/(k_{B}T)$ for the inverse temperature.
The self-energy $\Sigma$ is evaluated in the ladder approximation
and is conveniently expressed in terms of the T-matrix which is the
sum of ladder diagrams. In compact notation the equation for the many-body
T-matrix $T$ is $T=U+U\chi T,$ where $U$ is the momentum dependent
bare interaction potential and $\chi$ is the in-medium pair propagator.
The product $U\chi T$ is to be understood as a momenta convolutions.
The convolutions will in general give rise to angular integrals. However,
in relation to Feshbach resonances the important physics resides in
the energy dependence of the T-matrix to which we make a low energy
expansion. The angular integral can therefore be applied to the pair
propagator alone and the angular dependence vanishes from the T-matrix
equation. In general the T-matrix depends on the two incoming particle
momenta, $\vec{k}_{1},\vec{k}_{2}$ and two outgoing $\vec{k}_{1}^{\prime},\vec{k}_{2}^{\prime}.$
For the off-shell T-matrix there are in addition dependence on two
incoming $\omega_{1},\omega_{2}$ and two outgoing and $\omega_{1}^{\prime},\omega_{2}^{\prime}$
particle energies. It is convenient to pass to relative momenta $\vec{k},\vec{k}^{\prime}$
and energies $\omega,\omega^{\prime}$ and total momentum and energy,
$\vec{q},\Omega.$ For the on-shell total energy, $\Omega=\omega_{1}+\omega_{2}=\varepsilon_{k_{1}}+\varepsilon_{k_{2}}=2\varepsilon_{k}+\varepsilon_{q}^{B},$
with the free fermion dispersion relation, $\varepsilon_{k}=\hbar^{2}k^{2}/(2m),$
the relative momentum in center-of-mass frame is $k=|\vec{k}|,$ the
center-of-mass momentum is $q=|\vec{q}|$ and the free pair dispersion
is $\varepsilon_{q}^{B}=\hbar^{2}q^{2}/(2m_{B})$with $m_{B}=2m.$
The analytically continued, $i\Omega_{\nu}\to\Omega+i0^{+},$ in-medium
pair propagator is \begin{eqnarray}
\chi(\vec{k},\vec{q},\Omega) & = & \frac{1}{\beta}\sum_{\omega_{n}}G_{0}(\xi_{+},\Omega)G_{0}(\xi_{-},\Omega-\omega_{n}),\end{eqnarray}
 where $\xi_{\pm}=\varepsilon_{\pm\vec{k}+\vec{q}/2}-\mu.$ We tacitly
used the fact that the T-matrix is independent of the relative Matsubara
frequencies. Let the two-body limit of the T-matrix be denoted by
$T_{0}$ and the free two-particle propagator by $\chi_{0}.$ The
corresponding two-body limit of the many-body T-matrix equation then
becomes $T_{0}=U+U\chi_{0}T_{0}.$ The renormalization procedure consists
of re-expressing the T-matrix in terms of two-body quantities which
on the other hand can be expressed directly in terms of the scattering
phase shift of the two-particle wave function in relative coordinates.
Heuristically, we may define $L$ from the equation $U=(1-\chi_{0})T_{0}\equiv LT_{0}$
and therefore find $T=L^{-1}LT=L^{-1}(T-\chi_{0}T)=L^{-1}U(1+(\chi-\chi_{0})T).$
Hence, after eliminating the bare interaction the T-matrix equation
becomes 

\begin{equation}
T=T_{0}+T_{0}(\chi-\chi_{0})T.\label{eq:TMA}\end{equation}
The off-shell two-body T-matrix $T_{0}(k)$ is related to the two-particle
off-shell scattering amplitude $f(k)$, as $T_{0}(k)=-4\pi\hbar^{2}f(k)/m=-[g(k)+ik]^{-1},$
where we introduced the relative momentum of the two colliding atoms,
$k=\sqrt{mE}=\sqrt{m}\left(\Omega-\xi_{q}^{B}\right)^{1/2}$ and where
for small momenta the function $g(k)=k\cot\delta_{0}$ can be expanded
in relative energy $\sim\hbar^{2}k^{2}/m.$ Thus, $g(k)=-a_{F}^{-1}+r_{\mathrm{e}}k^{2}/2+\ldots,$
where $a_{F}$ is the scattering length of the fermions, and $r_{\mathrm{e}}$is
the effective range. The \textit{shape-independent approximation}
applied here consists of using only the two first expansion coefficients
which characterizes the potential by its depth and range, but is more
generally also a valid approximation for multichannel scattering problems.
For model potentials the shape dependence is reflected in the signs
and magnitudes of the higher order expansion coefficients. The effective
range has dimension of length and is for an attractive monotonic potential
at resonance, $a_{F}\to\pm\infty,$ the range of the inter-atomic
potential. For the resonance phenomena considered here the effective
range can be large and negative. Inserting the low energy expansion
of $g(k)$ into the scattering amplitude, $f(k)=k^{-1}e^{i\delta_{0}}\sin(\delta_{0})=(-a_{F}^{-1}+r_{\mathrm{e}}k^{2}/2-ik)^{-1},$
the two-body T-matrix becomes, $T_{0}(k)=-(4\pi a_{F}/m)(1-r_{\mathrm{e}}a_{F}k^{2}/2-ik)^{-1}.$
Adding and subtracting $1/2\varepsilon_{k^{\prime}}$ in Eq. \prettyref{eq:TMA}
and using the identity \begin{equation}
ik=\frac{1}{\mathcal{V}}\sum_{\vec{k}^{\prime}}\left(\chi_{0}(\vec{k}^{\prime},\vec{q},\Omega)-\frac{1}{2\varepsilon_{k^{\prime}}}\right),\end{equation}
 the free two-particle pair propagator, $\chi_{0},$ can be eliminated
from the problem. After inserting the expression for $ik$ into the
Eq. \prettyref{eq:TMA} the inverse effective interaction can be defined
as \begin{equation}
U_{\ast}^{-1}(k)=-\frac{m}{4\pi\hbar^{2}}\left(\frac{1}{a_{F}}-\frac{1}{2}r_{\mathrm{e}}k^{2}\right).\end{equation}
 By solving the T-matrix equation the expression for the inverse T-matrix
becomes \begin{equation}
T^{-1}(\vec{q},\Omega)=\left(U_{\ast}^{-1}(E)-\tilde{\chi}(\vec{q},\Omega)\right),\end{equation}
where the regularized nonzero temperature in-medium pair propagator
is, 

\begin{equation}
\tilde{\chi}(\vec{q},\Omega)=\frac{1}{\mathcal{V}}\sum_{\vec{k}^{\prime}}\left(\chi(\vec{k}^{\prime},\vec{q},\Omega)-\frac{1}{2\varepsilon_{k^{\prime}}}\right).\end{equation}
 The Matsubara sum in $\chi(\vec{k}^{\prime},\vec{q},\Omega)$ is
evaluated by turning the sum over Matsubara poles into a contour integral
using standard techniques \cite{Fetter1971}. The regularized pair
propagator $\tilde{\chi}$ then becomes\begin{eqnarray}
\tilde{\chi}(\vec{q},\Omega) & = & \frac{1}{\mathcal{V}}\sum_{\vec{k}^{\prime}}\frac{1-n_{F}(\xi_{+})-n_{F}(\xi_{-}-\Omega)}{\xi_{+}+\xi_{-}-\Omega}-\frac{1}{2\varepsilon_{\vec{k}^{\prime}}},\label{eq:pair-reg}\end{eqnarray}
where $n_{F}(\xi)=(e^{\beta\xi}+1)^{-1}$ is the Fermi-Dirac distribution
function. At a later stage we are going to make a low $q,\Omega$
expansion of $\tilde{\chi}$ and therefore keep the $\Omega$ in $n_{F}$
in the numerator \cite{Quintanilla2002}. For certain values of $\beta$
and $\mu$ the pair propagator $\tilde{\chi}$ can be calculated analytically,
but it is generally evaluated by numerical integration. The angular
dependence $x\equiv\cos\theta=\vec{k}^{\prime}\cdot\vec{q}/k^{\prime}q$
is absent in the denominator and the angular integration can be carried
out analytically in the numerator. The regularized pair propagator
thus reduces to a one dimensional integral over the magnitude of the
relative momentum. After renormalization and making the effective
range expansion the equation for the T-matrix is of the same form
as in the corresponding equation obtained from the bare contact potential
with the only difference being the energy dependence of the effective
interaction, $U_{\ast}\to U_{\ast}(E)$ where $E=\hbar^{2}k^{2}/m.$
Having obtained the expression for the T-matrix the self-energy becomes
\begin{equation}
\Sigma(\vec{k},\omega_{n})=\frac{1}{\beta\mathcal{V}}\sum_{\vec{q},\Omega_{\nu}}T(\vec{q},\Omega_{\nu})G_{0}(\vec{q}-\vec{k},\Omega_{\nu}-\omega_{n}),\label{eq:selfenergy}\end{equation}
where $\Omega_{\nu}=2\pi n/\beta$ for integer $\nu$ are the bosonic
Matsubara frequencies. 

To locate the transition temperature we apply the Thouless criterion
\cite{Thouless1960a} which states that the transition to the superfluid
state is found from the singular behavior of the T-matrix in the long
wavelength and low energy limit. At $T_{c}$ the Thouless criterion
is equivalent to the gap equation. For a fixed number of atoms and
at the transition temperature the crossover equations are then $T^{-1}(\vec{q},\Omega)=0$
for $q\to0$ and $\Omega\to0$ together with the equation for the
particle number conservation. In order to calculate the particle number
density $n=N/\mathcal{V}$ self-consistently we use the thermodynamic
relation, $N=-\partial_{\mu}\Omega,$ where $\Omega$ is the grand
canonical potential, $e^{-\beta\Omega}=\mathrm{Tr}\left[e^{-\beta K}\right],$
tracing over all degrees of freedom. At the transition the result
for the normal phase is still valid and the thermodynamic potential
can be found from the Luttinger-Ward expression for the grand potential
\begin{equation}
\Omega[G,\Sigma]=\Phi[G]-\frac{1}{\beta\mathcal{V}}\sum_{\vec{k},\omega_{n},\sigma}e^{i\Omega_{\nu}0^{+}}\left[\ln(-G_{0}^{-1}+\Sigma)+\Sigma G\right],\end{equation}
 where the physical dressed Green's function $G(\vec{k},\omega)$
for the atoms and the self-energy $\Sigma(\vec{k},\omega_{n})$ are
found from the stationary points of $\Omega$ with respect to variations
of both $G$ and $\Sigma.$ Here, $\delta\Omega/\delta\Sigma=0$ yields
the Dyson equation Eq. \prettyref{eq:dyson} and $\delta\Omega/\delta G=0$
links the self-energy, $\Sigma,$ to the functional $\Phi$ which
is the generating functional for the skeleton graphs and is here constructed
in order to generate the ladder approximation by functional differentiation,
$\Sigma=\frac{1}{2}\delta\Phi/\delta G$ \cite{Baym1962}. In the
ladder approximation\begin{equation}
\Phi=\frac{1}{\beta\mathcal{V}}\sum_{\vec{q},\Omega_{\nu}}\ln(1-U_{\ast}\chi).\end{equation}
 A self-consistent treatment of the zero effective range crossover
problem has been presented in \cite{Haussmann1993,Haussmann1994,Haussmann2000}.
We have employed a non-selfconsistent approach where the dressed Green's
functions in the pair propagator and the self-energy has been replaced
by $G_{0}$ and which is equivalent to the functional integral approach
of \cite{Melo1993}. In this approximation the grand canonical potential
takes the form, $\Omega=\Omega_{0}^{F}+\delta\Omega,$ with $\Omega_{0}^{F}=\sum_{\vec{k},\omega_{n},\sigma}e^{i\omega_{n}0^{+}}\ln G_{0}(\vec{k},\omega_{n})/\beta\mathcal{V}$
and $\delta\Omega=\sum_{\vec{q},\Omega_{\nu}}e^{i\Omega_{\nu}0^{+}}\ln T(\vec{q},\Omega_{\nu})/\beta\mathcal{V}$
\cite{Thouless1960a}.

\section{Transition Temperature\label{sec:transition-temperature} }

In a BCS superfluid the dispersion relation for the quasiparticle
excitations is, $E_{\vec{k}}=(\xi_{\vec{k}}^{2}+\Delta^{2})^{1/2},$
and which displays an energy gap, $\Delta\sim U\langle\psi\psi\rangle,$
with $\langle\psi\psi\rangle$ being a anomalous thermal correlator.
Due to the energy gap in the single particle spectrum thermodynamic
and transport quantities, such as the heat capacity and thermal conductivity,
may become exponentially suppressed in the superfluid phase and may
therefore be used as thermodynamic signatures of the phase transition.
Apart from such quantities being difficult to observe directly in
a trapped atom gas there are in a strongly interacting Fermi gas additional
contributions to the quasiparticle gap which may further complicate
the identification of the superfluid transition from such observables.
A simple way to see the appearance of such a non-BCS gap contribution
at zero temperature is to note that the gap in the quasiparticle spectrum
is the minimum of $E_{\vec{k}}$ over all $\vec{k}$'s. In the BCS
limit the chemical potential is positive and approximately the Fermi
energy, and the gap is $\Delta$ for $k=k_{F}.$ By contrast when
the chemical potential becomes negative on the BEC side the minimum
is shifted to $k=0$ for the gap $(\mu^{2}+\Delta^{2})^{1/2}$ \cite{Leggett1980}.
The appearance of $\mu$ in $E_{k}$ is related to the molecular degrees
of freedom which also contribute to the gap at higher temperatures.

\subsection{Scattering length only \label{sub:Scattering-length-only}}

\begin{figure}
\begin{center}\includegraphics[%
  width=8cm]{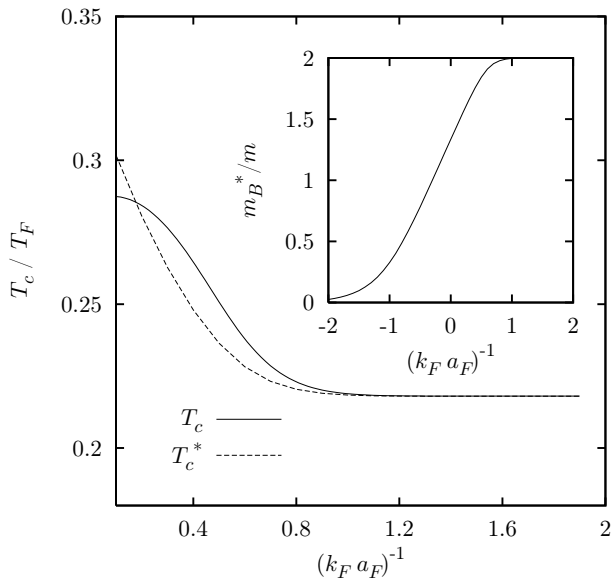}\end{center}

\caption{The transition temperature for $r_{\mathrm{e}}=0$ on the BEC side
of the crossover is plotted as a function of the inverse coupling
constant from the full calculation (full curve) and the corresponding
result (dashed curve) obtained from the expression for the transition
temperature for the non-interacting BEC with the numerically calculated
effective boson mass. The inset shows the crossover behavior of the
effective boson mass in units of the fermion mass in the whole crossover
region. \label{cap:bec-Tc}}
\end{figure}
 Let us initially consider the zero range limit, $r_{\mathrm{e}}=0,$
which results in the standard BCS to BEC crossover of a Fermi gas
with contact interaction \cite{Nozieres1985,Melo1993}. The transition
temperature $T_{c}=1/\beta_{c}$ and the chemical potential $\mu_{c}$
are at the transition determined from the Thouless criterion, $\lim_{q,\Omega\to0}T^{-1}(\vec{q},\Omega)=0,$
together with the number density equation, $n=n_{0}^{F}(\beta,\mu)+\delta n(\beta,\mu)$,
where the fermion contribution is $n_{0}^{F}(\beta,\mu)=\sum_{\vec{k},\sigma}n_{F}(\xi_{k})/\mathcal{V}.$
The fluctuation term $\delta n$ derived from the ladder self-energy
contribution to the thermodynamic potential is $\delta n(\beta,\mu)=-\sum_{\vec{q}}\sum_{\Omega_{\nu}}\partial_{\mu}\ln T(\vec{q},\Omega_{\nu})/\beta\mathcal{V}.$
To simplify the calculation of the bosonic number density we follow
\cite{Drechsler1992,Melo1993,Stintzing1997,Chen1998,Quintanilla2002}
and make a low energy and long wavelength expansion of the inverse
T-matrix \begin{eqnarray}
T^{-1}(\vec{q},\Omega) & = & Z(\Omega-\varepsilon_{q}^{B}-\mu_{B}),\label{eq:boson-prop}\end{eqnarray}
 where $i\Omega_{\nu}\to\Omega$ before frequency expansion, the molecular
dispersion relation is $\varepsilon_{q}^{B}=\hbar^{2}\vec{q}^{2}/2m_{B}^{\ast}$
with $m_{B}^{\ast}$ being the renormalized boson mass, the wave function
renormalization constant $Z=Z(\beta,\mu),$ the effective boson mass
$m_{B}^{\ast}=m_{B}^{\ast}(\beta,\mu)$ and the bosonic chemical potential
$\mu_{B}=\mu_{B}(\beta,\mu)$ are found from the low order energy
and momentum expansion coefficients of $\tilde{\chi}$ together with
$U_{\ast}^{-1}(k).$ Re-expressing the frequency integration as a
standard sum over bosonic Matsubara frequencies $\delta n$ becomes
\begin{eqnarray}
\delta n(\beta,\mu) & = & \frac{1}{\mathcal{V}}\sum_{\vec{q}}n_{B}(\xi_{q}^{B})(-\partial_{\mu}\xi_{q}^{B}),\end{eqnarray}
 where $\xi_{q}^{B}=\varepsilon_{q}^{B}-\mu_{B},$ and $n_{B}(\xi)=(e^{\beta\xi}-1)^{-1}$
is the Bose-Einstein distribution function. 

Before we turn to the results of the numerical calculations we note
that the results for the uniform case with a contact potential are
well-known: In the weak coupling limit as $(k_{F}a_{F})^{-1}\to-\infty$
the Fermi gas is expected to make a transition at the transition temperature
$T_{c}=(8e^{\gamma-2}/\pi)T_{F}e^{-\pi/2k_{F}|a_{F}|}$ to a pair-correlated
BCS-state, where the Fermi temperature is, $T_{F}=E_{F}/k_{B},$ and
$\gamma=0.57721\ldots$ is the Euler-Mascheroni constant. For a uniform
noninteracting two-component Fermi gas with the Fermi energy $E_{F}$
the corresponding Fermi wave vector is $k_{F}=(2mE_{F}/\hbar^{2})^{1/2},$
and the particle density is $n=k_{F}^{3}/3\pi^{2}$. As the inverse
scattering length is increased strong coupling effects appear and
when $(k_{F}a_{F})^{-1}$ passes through zero from below a two-body
bound (molecular) state appears, and the fermionic states hybridize
with the molecular states. When the inverse scattering length is increased
further $(k_{F}a_{F})^{-1}\to\infty$ the Fermi surface smears out
(which can be seen by inspection of the momentum distribution) and
a molecular Bose gas is formed which undergoes Bose-Einstein condensation
at the transition temperature $T_{BEC}=(\hbar^{2}/2mk_{B})2\pi\zeta(3/2)^{-2/3}(n/2)^{2/3}\approx0.218T_{F}.$
Between the two limits the system smoothly crosses from BCS-like behavior
to BEC-like behavior in the region, $-1\lesssim(k_{F}a_{F})^{-1}\lesssim1$.
In the BEC limit where the fermion contribution vanishes $\delta n$
reduces to twice the boson density obtained from the Bose-Einstein
distribution, $\delta n(\beta,\mu)=2\sum_{\vec{q}}n_{B}(\xi_{q}^{b})/\mathcal{V},$
for bosons with the mass twice that of the atoms, $m_{B}^{\ast}=m_{B}=2m.$ 

The problem of numerically solving the pair of equations $n=n_{0}+\delta n$
and $T^{-1}(\vec{0},0)=0$ for $\beta_{c}$ and $\mu_{c}$ was solved
iteratively using multidimensional root solvers implemented in \cite{Galassi2003}.
At each iteration the momentum integrals appearing in $Z,m_{B}^{\ast},\delta n,\mu_{B}$
and in the $\mu$ derivatives appearing in the number equation were
numerically integrated using appropriate adaptive quadrature algorithms
\cite{Piessens1983} also reimplemented in \cite{Galassi2003}. The
numerical results for the zero effective range crossover are presented
in Fig. \ref{cap:Tc-BCS-to-BEC} (full curve) and show a smooth crossover
with characteristic $T_{c}$ peak near the zero of the chemical potential
and qualitatively in agreement with previous calculations \cite{Nozieres1985,Melo1993}. 

For a weakly interacting Bose gas there are a number of corrections
due to critical fluctuations close to $T_{c}$ with the leading term
being linear and positive in the bosonic gas parameter $\propto k_{F}a_{F}$
and indicating a peak in $T_{c}.$ Mean field corrections to the transition
temperature for a weakly interacting Bose gas with nonzero effective
range give rise to a leading correction term which is negative and
cubic in the gas parameter and thus indicates absence of peak in $T_{c}.$
However, the non-selfconsistent Gaussian model presented here includes
none of the above mentioned boson-boson effects and the correction
to $T_{c}$ observed on the BEC side here is due to an externally
induced change of the molecular binding energy $E_{0}.$ When $(k_{F}a_{F})^{-1}$
is lowered by increasing the strength of the effective interaction
the binding energy goes down and the molecules dissociate more easily
at a fixed temperature and in addition the fermion states become populated.
The coupling to the fermions also gives rise to a renormalized boson
mass, $m_{B}^{\ast}<m_{B},$ where the bare boson mass is $m_{B}=2m.$
Therefore the ideal Bose gas condensation temperature is shifted to
higher values, $T_{c}=T_{BEC}+\delta T_{BEC}$ where $\delta T_{BEC}/T_{BEC}\propto|\delta m_{B}|/m_{B}$
for decreasing effective mass, $\delta m_{B}=m_{B}^{\ast}-m_{B}$
as $(k_{F}a_{F})^{-1}$ increases. In Figure \ref{cap:bec-Tc} the
transition temperature in the BEC limit is plotted as a function of
the $(k_{F}a_{F})^{-1}$ using the numerically calculated $T_{c}$
(full curve) and using the ideal Bose gas expression for the transition
temperature with $T_{c}^{\ast}\equiv T_{BEC}^{\ast}\propto1/m_{B}^{\ast}$
using the numerically calculated effective mass $m_{B}^{\ast}.$ The
inset in Figure \ref{cap:bec-Tc} shows the effective mass $m_{B}^{\ast}$
identified from the asymptotic behavior of the T-matrix, see Eq. \prettyref{eq:boson-prop}.
The effective mass becomes as expected twice the fermion mass in the
BEC limit, but no attempt was made to decompose $\delta n$ into partial
contributions. Thus, the effective boson mass becomes a hybrid object
in the crossover region without clear physical interpretation as a
single particle mass.

\subsection{Effective range expansion \label{sub:Effective-range-expansion} }

\begin{figure}
\begin{center}\includegraphics[%
  width=8cm]{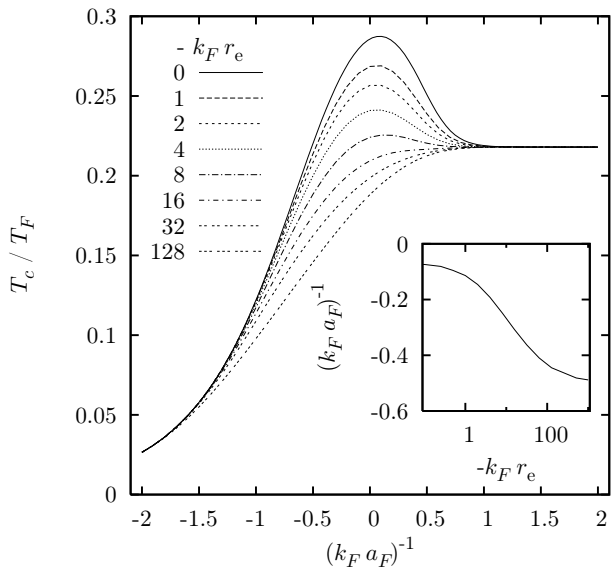}\end{center}

\caption{The BCS-BEC crossover transition temperature of the homogeneous Fermi
gas as a function of the inverse coupling constant, $-2\leq(k_{F}a_{F})^{-1}\leq2$,
for several values of the effective range parameter $k_{F}r_{\mathrm{e}}.$
The zero range result (full curve) corresponds to the usual $T_{c}$
curve for the crossover. For increasing values of the $-k_{F}r_{e}$
the transition temperature is suppressed in the crossover region.
In the inset we characterize the small effective range to large effective
range crossover by plotting the half-density crossing point in $(k_{F}a_{F})^{-1}$
as a function of the $-k_{F}r_{e}.$ The curve shows a shift in the
position of the crossing to lower couplings for increasing $-r_{e}.$
\label{cap:Tc-BCS-to-BEC}}
\end{figure}
 As indicated in the introduction narrow Feshbach resonances can be
mimicked by an interaction potential with large and negative effective
range. To account for such a situation the zero effective range contact
potential must be replaced by a potential with nonzero range. However,
for monotonic and attractive potentials the effective range is at
resonance the averaged microscopic range of the potential and thus
positive \cite{Blatt1949}. For potential models the effective range
can be related to the structure of the interaction potential and can
be negative in the presence of repulsive cores or barriers. In the
latter case considered here the negative effective range is related
to the resonance physics of potentials with barriers. For a potential
with a barrier quasi-bound states may appear which have a finite lifetime
and thus decay into the continuum states. In the zero temperature
calculation \cite{Palo2004,Diener2004a} it was suggested to mimic
the presence Feshbach resonances by using a two-body square well with
a square barrier model potential in place of the contact potential.
In contrast we use the renormalized crossover equations of Section
\ref{sec:intro} and the numerics of subsection \ref{sub:Scattering-length-only}
to examine the effect of the energy dependent effective interaction
\begin{eqnarray}
U_{\ast}(k) & = & -\frac{4\pi a_{F}}{m}\left(1-\frac{1}{2}r_{\mathrm{e}}a_{F}k^{2}\right)^{-1}\end{eqnarray}
 on the transition temperature for values of the effective range $-k_{F}r_{\mathrm{e}}$
ranging from small to large and positive. 

In Figure \ref{cap:Tc-BCS-to-BEC} the numerical results for the transition
temperature for several values of $-k_{F}r_{\mathrm{e}}=1,2,4,8,16,32,128$
are plotted as a function of the $(k_{F}a_{F})^{-1}$ together with
the zero range $r_{\mathrm{e}}=0$ result (full curve). For a broad
resonance with vanishing $k_{F}r_{\mathrm{e}}$ the usual zero effective
range result is obtained. In the crossover region the transition temperature
is sensitive to the effective range and the peak in $T_{c}$ becomes
suppressed for increasing values of $-k_{F}r_{\mathrm{e}}.$ In the
BCS (BEC) limit a negative $r_{\mathrm{e}}$ shifts the effective
interaction to lower (higher) values corresponding to a lower transition
temperature which is consistent with the decrease in $T_{c}$ for
increasing $-r_{\mathrm{e}}$ at fixed $a_{F}.$ Furthermore, it is
seen that the energy dependence of the effective interaction is only
important in the crossover region for $-k_{F}r_{\mathrm{e}}\gtrsim1.$
The small value of the parameter $(k_{F}a_{F})(k_{F}r_{\mathrm{e}})$
the energy dependence of $U_{\ast}$ is suppressed and therefore both
the asymptotic BCS and BEC limits are recovered even for large $k_{F}r_{\mathrm{e}}.$ 

It is tempting to compare our result to the corresponding calculation
of $T_{c}$ for the atom-molecule Hamiltonian. In \cite{Ohashi2002a,Ohashi2003a}
the transition temperature was calculated as a function of the detuning
$\nu$ for a homogeneous Fermi gas both for weak inter-channel coupling
$g_{BF}=0.6E_{F}$ and for strong coupling $g_{BF}=20E_{F}.$ In addition
to an explicit contribution to the number equation arising from the
bosonic degrees of freedom, the effective interaction appearing in
the Thouless criterion also includes the effect of repeated fermion-boson
scattering in addition to the fermion-fermion scattering, $U_{\ast}(\vec{q},\Omega_{\nu})=U_{FF}-g_{BF}^{2}D_{0}(\vec{q},\Omega_{\nu}),$
where $U_{FF}$ is the inter-fermion coupling strength, $g_{BF}$
is the boson-fermion coupling strength, $D_{0}^{=1}(\vec{q},\Omega_{\nu})=(i\Omega_{\nu}-\xi_{q}^{B})^{-1}$
is the free molecule propagator, $\xi_{q}^{B}=\hbar^{2}q^{2}/(2m_{B})-2(\mu-\nu)$
is the bare molecule dispersion relation. For small couplings corresponding
to a large and negative effective range $T_{c}$ was found to increase
monotonically from the BCS to the BEC limit. In the opposite limit
of large couplings corresponding to a small effective range a pronounced
peak close to threshold appears. These results are consistent with
the trends found in the present calculation. It is also interesting
to compare the result to the transition temperature obtained for a
mixture of fermions and long-lived bosons (narrow resonance case)
with the total density constrained by total fermion number conservation,
and with $m_{B}=2m.$ By decomposing the total density into three
contributions, $n=n_{0}^{F}+2n_{0}^{B}+\delta\tilde{n,}$ and by neglecting
$\delta\tilde{n}$ when solving the crossover equations, a monotonically
increasing $T_{c}$ is found which appears to be the $T_{c}$ curve
in the large $-r_{\mathrm{e}}$ limit. The effective range induce
a cutoff in the momentum integrations and thus reduces the influence
of high momentum modes. This decreases the effect of the fluctuations
and leads to a suppression of $T_{c}$ which is again consistent with
the interpretation of the large $T_{c}$ peak being due to Gaussian
fluctutations. To characterize the 'crossover' from small $r_{\mathrm{e}}$
behavior to large $r_{\mathrm{e}}$ behavior we notice that at a fixed
$k_{F}a_{F}$ the fermion contribution to the total density becomes
suppressed for increasing $-r_{\mathrm{e}}$ due to decrease in the
chemical potential in the crossover region. Therefore, the half-density
crossing point of the partial densities, $n_{0}^{F}=\delta n=\frac{1}{2},$
is shifted to the BCS side. The inset in Fig. \ref{cap:Tc-BCS-to-BEC}
shows a semilogarithmic plot of the half-density crossing point as
a function of $k_{F}r_{\mathrm{e}}.$ As value of the effective range
is increased the crossing point shifts from $(k_{F}a_{F})^{-1}\approx-0.07$
for $r_{\mathrm{e}}=0$ to $(k_{F}a_{F})^{-1}\approx-0.5$ in the
large $-r_{\mathrm{e}}$ limit where the dependence on the effective
range is expected to drop out. Therefore, the transition temperature
appear to have a lower bound at fixed coupling for large $-r_{\mathrm{e}}$
when neglecting high order terms in the expansion of $k\cot\delta.$ 

It is to be noted that the current calculation does not account for
i) the induced interactions due to vertex corrections which reduce
$T_{c}$ by a factor of $2.2$ in the BCS limit \cite{Gorkov1961}.
ii) the residual boson-boson and boson-fermion interactions which
are characterized by the boson-boson scattering length, $a_{B},$
and the boson-fermion scattering length, $a_{BF},$ on the BEC side
as recently found from solving three and four-body problems \cite{Petrov2004b}.
Let us finally mention that to the extent that many-body physics is
important to the crossover problem it must be stressed that many theoretical
issues, e.g., non-zero effective range calculations, renormalization
of coupling constants, inclusion of higher order diagrammatic contributions,
and generalization to pairing states in higher angular momentum channels,
are easier to handle within the single channel potential model.

\section{Summary}

We presented the renormalized T-matrix equation which allowed the
non-zero effective range BCS-BEC crossover equations at the transition
to be formulated directly in terms of two-body scattering properties,
the scattering length and the effective range. By matching the scattering
properties of the generalized single-channel BCS-BEC crossover model
with the two-channel Feshbach resonance atom-molecule model it was
argued that the internal energy scale of a resonance translates into
to the effective range of the potential model. Broad resonances correspond
to a small effective range and for narrow resonances the effective
range become large and negative. By using the NS crossover equations
with the self-energy derived from the ladder approximation we numerically
calculated the superfluid transition temperature of the generalized
BCS-BEC crossover model as a function of the two-body scattering length
and found $T_{c}$ to be suppressed in the crossover region for increasing
values of $-k_{F}r_{\mathrm{e}}.$

\begin{acknowledgments}
It is a pleasure to acknowledge helpful discussions with Chris Pethick. 
\end{acknowledgments}
\bibliographystyle{apsrev}
\bibliography{/home/melwyn/work/strong_coupling/papers/eff_range/bec,/home/melwyn/work/strong_coupling/papers/eff_range/mel,/home/melwyn/work/strong_coupling/papers/eff_range/sqwb}

\end{document}